\documentclass[%
 reprint,
superscriptaddress,
 amsmath,amssymb,
 aps,
pra,
]{revtex4-2}

\usepackage{graphicx}
\usepackage{dcolumn}
\usepackage{bm}
\usepackage{hyperref}
\usepackage[separate-uncertainty=true]{siunitx}

\begin{document}


\title{Fracture metamaterials with on-demand crack paths enabled by bending}

\author{Lucie Domino}
\affiliation{Institute of Physics, Universiteit van Amsterdam, Science Park 904, 1098 XH Amsterdam, The Netherlands}
\affiliation{Université libre de Bruxelles (ULB), Nonlinear Physical Chemistry Unit, CP 231, 1050 Bruxelles, Belgium}

\author{Mariam Beaure d'Augères} 
\affiliation{Institute of Physics, Universiteit van Amsterdam, Science Park 904, 1098 XH Amsterdam, The Netherlands}

\author{Jian Zhang}
\affiliation{Department of Mechanical Engineering, Eindhoven University of Technology, Den Dolech 2, 5612 AZ Eindhoven, The Netherlands}
\affiliation{Faculty of Mechanical Engineering, Delft University of Technology, Mekelweg 2, 2628 CD Delft, The Netherlands}

\author{Shahram Janbaz}
\affiliation{Institute of Physics, Universiteit van Amsterdam, Science Park 904, 1098 XH Amsterdam, The Netherlands}

\author{Alejandro M.~Arag\'{o}n}
\affiliation{Faculty of Mechanical Engineering, Delft University of Technology, Mekelweg 2, 2628 CD Delft, The Netherlands}

\author{Corentin Coulais}
\affiliation{Institute of Physics, Universiteit van Amsterdam, Science Park 904, 1098 XH Amsterdam, The Netherlands}

\date{\today}

\begin{abstract}
In many scenarios---when we bite food or during a crash---fracture is inevitable. Finding solutions to steer fracture to mitigate its impact or turn it into a purposeful functionality, is therefore crucial. Strategies using composites, changes in chemical composition or crystal orientation, have proven to be very efficient, but the crack path control remains limited and has not been achieved in load-bearing structures. Here, we introduce fracture metamaterials consisting of slender elements whose bending enables large elastic deformation as fracture propagates. This interplay between bending and fracture enables tunable energy dissipation and the design of on-demand crack paths of arbitrary complexity. To this end, we use topology optimisation to create unit cells with anisotropic fracture energy, which we then tile up to realize fracture metamaterials with uniform density that we 3D-print. The thin ligaments that constitute the unit cells confer them a strikingly distinct response in tension and shear, and we show that by controlling the orientation and layout of the unit cells the sequential progress of the crack can be controlled, making the fracture path arbitrarily tortuous. This tortuosity increases the energy dissipation of the metamaterial without changing its stiffness. Using bespoke arrangements of unit cells, metamaterials can have on-demand fracture paths of arbitrary complexity. Our findings bring a new perspective on inelastic deformations in mechanical metamaterials, with potential applications in areas as diverse as the food industry, structural design, and for shock and impact damping.

\end{abstract}

\maketitle



\begin{figure}[t!]
\centering
  \includegraphics[width=1\columnwidth,clip,trim=0cm 0cm 0cm 0cm]{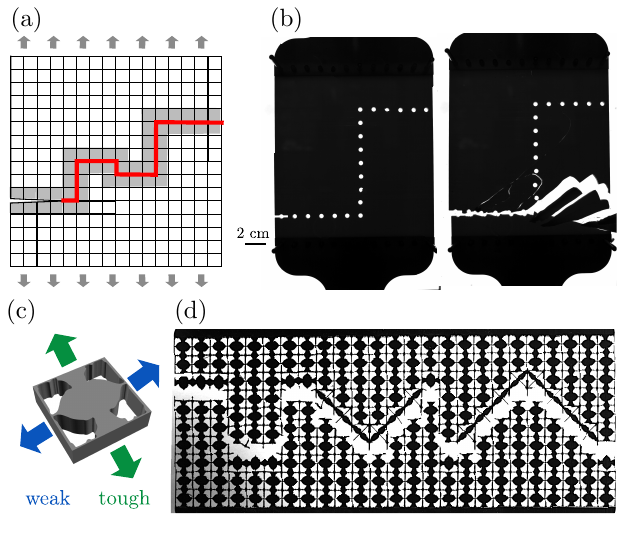}
 \caption{\textbf{Controlling a fracture path with unit cells with giant fracture toughness anisotropy.} 
 (a) In this article, we address the challenge of how to achieve designer cracks of arbitrary complexity. 
 (b) Naive example of a step-shaped crack path made of holes in a homogeneous medium. 
 (c) Unit cell with giant anisotropic fracture toughness, $J_1/J_2 = 15$. 
 (d) Fracture metamaterial with a crack forming the letter ``U'', ``V'' and `` A'', see also Supplementary Movie S1.}
 \label{fig:1}
\end{figure}

From carving flints to cracking eggs open, fracture is a phenomenon that has always been put to good use by humankind. The traditional approach is to choose precisely the points of application of the load to control how the material breaks, as is done in thin films for example~\cite{liu_recent_2022, roman_fracture_2013}. 
An alternative approach is to tailor the material instead. For instance, highly anisotropic materials such as nacre~\cite{barthelat_mechanical_2006, barthelat_mechanics_2007}, wood~\cite{fruhmann_fracture_2002,cave_anisotropic_1968,katz_anisotropic_2008}, or cuttlefish bone~\cite{birchall_architecture_1983} guide cracks along one direction---they are extremely tough if one tries to break them in the other. Inspired by natural materials, scientists have devised materials with increased fracture toughness~\cite{dimas_modeling_2014, tang_nanostructured_2003, mirkhalaf_overcoming_2014, espinosa_merger_2009}. One can then wonder whether artificial metamaterials could be designed to achieve even finer control over their fracture properties, such as cracks of arbitrary complexity (Fig.~\hyperref[fig:1]{\ref{fig:1}a}) or with tunable energy dissipation.

\begin{figure*}[t!]
 \begin{center}
  \includegraphics[width=2\columnwidth,clip,trim=0cm 0cm 0cm 0cm]{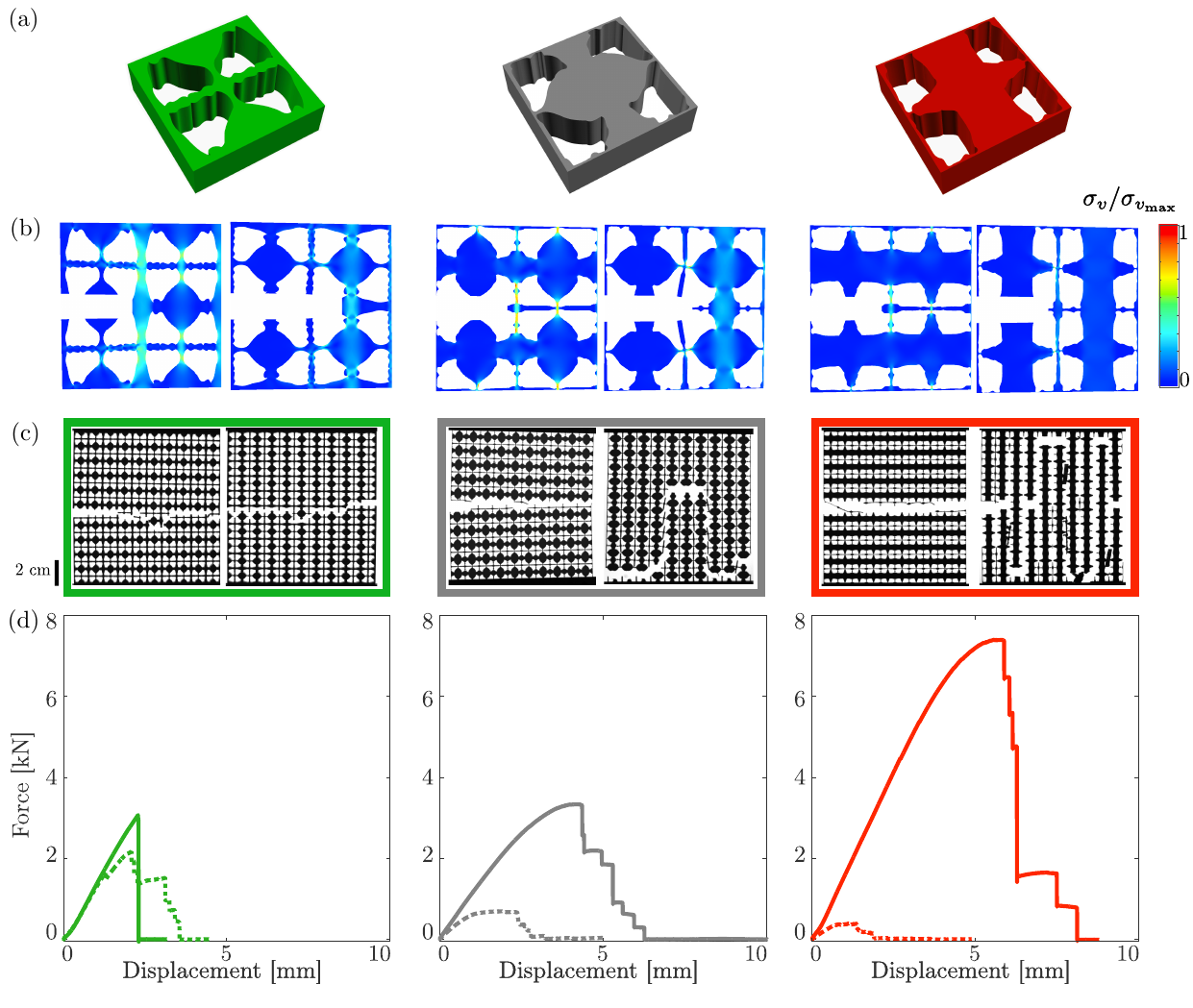}
 \end{center}
 \caption{\textbf{Larger anisotropy in the energy release rate increases both the crack length and the crack path complexity.} 
 (a) Three of the unit cells considered with anisotropy ratio $J_1/J_2=1.7$ (left), $J_1/J_2=15$ (middle), and $J_1/J_2=45$ (right).
 (b) Normalised von Mises stress $\sigma_v$ in the 4 central unit cells of a 10x10 sample under uniaxial tension, from finite element numerical simulation (see Methods section for details).
 (c) Crack paths obtained in uniform samples, with horizontal (left) or vertical (right) unit cells, see also Supplementary Movie S2.
 (d) Measured force-displacement curves for the horizontal (dashed line) and vertical (solid line) samples.
 }
 \label{fig:2}
\end{figure*}

Many promising approaches to steer cracks have been reported in the literature, such as designing specific notches in thin films~\cite{kim_cracking-assisted_2015}, patterning a sheet with cuts~\cite{hwang_metamaterial_2023} or designing complex adhesion with a substrate~\cite{xia_toughening_2012}. Using anisotropy is another common route to avoid mode I fracture in natural and synthetic composites~\cite{barthelat_architectured_2015, tang_nanostructured_2003, benedetti_architected_2021, martin_designing_2015}, in particular the use of microfibers with a selected and/or varying orientation~\cite{mesgarnejad_crack_2020, martin_designing_2015}. Yet in all those examples the crack path is ill-controlled (Fig.~\ref{fig:1}b). On-demand crack paths have been instead achieved using either local change of density~\cite{manno_engineering_2019}, stiffness~\cite{gao_crack_2020}, asymmetry~\cite{kaushik_experimental_2014, widstrand_stress_2023, brodnik_fracture_2021}, crystal orientation in thin films \cite{nam_patterning_2012}, and curvature~\cite{mitchell_fracture_2020, mitchell_fracture_2017}. Such materials however are not load-bearing, which limits their potential applications. Alternative strategies consist in tuning the rigidity of the material to switch from a brittle crack to a diffuse mode of failure~\cite{driscoll_role_2016}, or changing the connectivity of a network of meso-structured tiles~\cite{magrini_control_2023}, but in both examples this comes at the cost of predictability of the crack path.

\begin{figure*}[t!]
 \begin{center}
  \includegraphics[width=2\columnwidth,clip,trim=0cm 0cm 0cm 0cm]{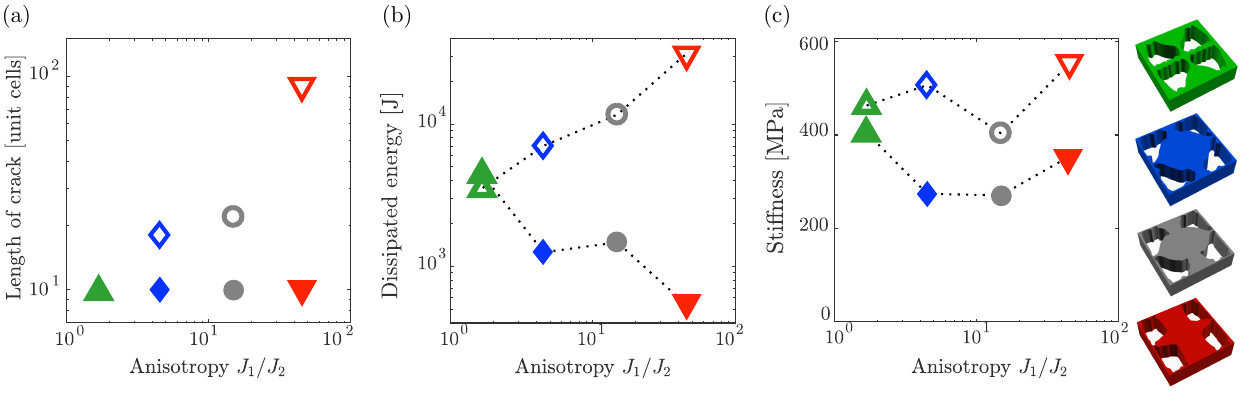}
 \end{center}
 \caption{\textbf{Role of anisotropy on crack length and mechanical properties.}
 Length of the crack measured in numbers of broken unit cells (a), dissipated energy (b) and stiffness (c) as a function of the fracture anisotropy ratio $J_1/J2$ of the unit cell (that takes values 1.7 (green upward triangles), 4.5 (blue diamonds), 15 (gray circles), 45 (red downward triangles)), measured in the 10 $\times$ 10 samples presented in figure 2c, made with either horizontal cells (full symbols) or vertical cells (empty symbols). Error bars are smaller than the symbols size, and the dotted lines are a guide for the eye.
 }
 \label{fig:3}
\end{figure*}

To circumvent these limitations, we propose a scale-free strategy to achieve on-demand crack paths in metamaterials created from virtually any brittle base material. Rather than optimising the whole material for a given crack path, which is unattainable with current optimisation techniques, we create metamaterials whose building block is designed to have specific fracture properties, namely extreme anisotropy in the critical energy release rate (Fig.~\hyperref[fig:1]{\ref{fig:1}c}). These are tiled to make metamaterials that, while keeping the volume of base material at a fixed ratio, break along on-demand crack paths (Fig.~\hyperref[fig:1]{\ref{fig:1}d}). We demonstrate the design strategy on 2D unit cells and several cases of emblematic fracture paths. Our findings illustrate a new fracture mechanism, where the least strong elements bend before breaking. We demonstrate that it is the very presence of these large bending deformations that enables the crack to align parallel to the loading direction and hence to follow an arbitrarily complex path. Our findings establish the use of slender elements within metamaterials as a distinctive avenue for the control of fracture.

\section*{Unit cells with anisotropic fracture}
We start by designing unit cells with highly anisotropic fracture behaviour. These are designed to exhibit very different critical energy release rate depending on the loading direction~\cite[\& references therein]{zhang_tailoring_2022, souto_edible_2022}. In other words, they are very weak if loaded in one direction but very tough if loaded in the perpendicular direction (Fig.~\hyperref[fig:1]{\ref{fig:1}c}). The unit cells are obtained by means of topology optimisation: starting from a square computational domain, the optimisation algorithm iteratively optimises the placement and removal of material to maximize the anisotropy in energy release rates for two loading directions~\cite{zhang_tailoring_2022} (see \hyperref[sec:methods]{Materials and Methods}). Notably, we also set a constraint on the volume of material, for instance at 50\%. Fracture anisotropy is quantified by the ratio $J_1/J_2$, where $J_i$ denotes an aggregation of energy release rates computed throughout the unit cell's boundary for the $i$th loading case. This quantity can be tuned, with values ranging from $J_1/J_2=1$ to $J_1/J_2=45$. We then tile the square unit cells to create a metamaterial with the required fracture properties. The question we address in this paper is how to choose the unit cells to create given fracture properties, such as a certain crack path like the one shown in Fig.~\hyperref[fig:1]{\ref{fig:1}d}, see also Supplementary Movie S1.

\section*{Tortuous cracks in uniform metamaterials} 
Let us first consider uniform samples, that is to say samples made with the same unit cell, oriented either horizontally or vertically. We use four different types of unit cells, three of which are shown in Fig.~\hyperref[fig:2]{\ref{fig:2}a}. We first consider a 100-unit cells metamaterial ($10 \times 10$ arrangement) and we perform a numerical analysis for a uniaxial static mode-I loading, \emph{i.e.} vertical direction in Fig.~\hyperref[fig:2]{\ref{fig:2}b}, where only the 4 central cells are shown (see \hyperref[sec:methods]{Materials and Methods}). The results of this simulation reveal that the stress (or strain) field in front of a notch has the highest values in the smallest ligaments, even if they are not in front of the notch. This means that for orientations of the unit cells where the weak direction is parallel to that of the loading, the crack would tend to avoid mode I fracture (perpendicular to the loading direction) and preferentially propagate parallel to the loading direction.

\begin{figure*}[t!]
 \begin{center}
  \includegraphics[width=2\columnwidth,clip,trim=0cm 0cm 0cm 0cm]{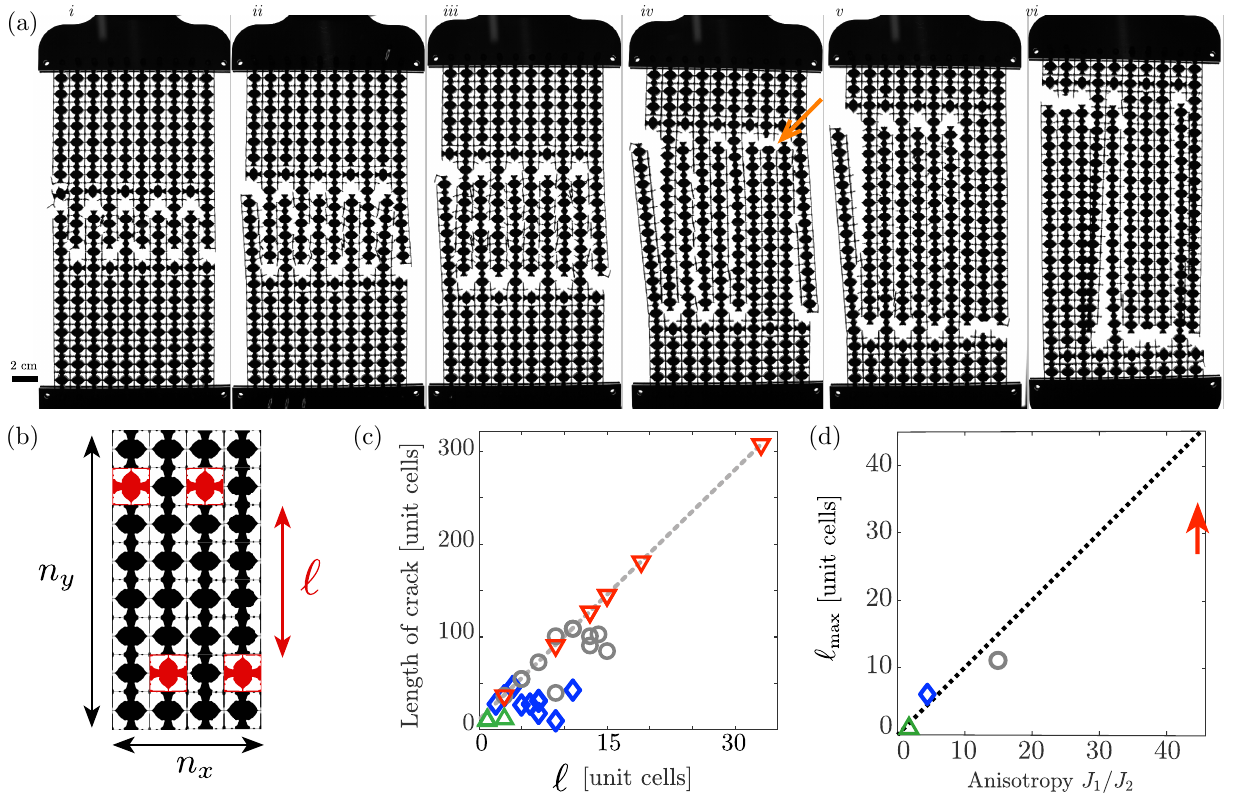}
 \end{center}
 \caption{\textbf{Anisotropy enhances designer cracks.} (a) Combs with an amplitude $\ell$ increasing from $\ell=3$ (left) to $\ell= 15$ (right), for $n_x = 10$, $n_y=19$, using the unit cell with fracture anisotropy ratio $J_1/J_2 = 15$. 
 The arrow in figure \textit{iv} denotes a horizontal shortcut taken by the crack, see also Supplementary Movie S3.
 (b) Sketch of the design principle used: $\ell$ is the number of unit cells between between two horizontally-tough cells in adjacent columns, and $n_x$ and $n_y$ are the number of unit cells in the specimen in the horizontal and vertical directions, respectively. 
 (c) Measured crack length as a function of the amplitude $\ell$ of the programmed crack path, for unit cells with a fracture anisotropy ratio of $1.7$ (green upward triangles), $4.5$ (blue diamonds), $15$ (dark circles), and $45$ (red downward triangles)). The uncertainty of the measured length is smaller than the size of the symbol. The dashed line represent the prediction $\ell \times (n_y-1) + n_x$, which corresponds to the length of the programmed crack. 
 (d) Maximum crack amplitude $\ell_{\max}$ achievable as a function of the fracture anisotropy ratio of the unit cells. The dashed line corresponds to the prediction $\ell_{\max} = J_1/J_2$.
}
 \label{fig:4}
\end{figure*}

We now perform experiments on samples composed of $10\times 10$ unit cells, that we fabricate using 3D printing (see \hyperref[sec:methods]{Materials and Methods}). These samples are then put under uniaxial tension in the vertical direction in a tensile testing machine (see \hyperref[sec:methods]{Materials and Methods} and Supplementary Information for a calibration of the constitutive material). We show in Fig.~\hyperref[fig:2]{\ref{fig:2}c} specimens where the unit cell's toughest direction is horizontal (left images) or vertical (right images). All the samples have a notch on the left to make sure the crack always starts at the same location\footnote{One exception is the sample with $J_1/J_2=15$ in the vertical orientation, which has been tested without notch.}. The fracture path for samples with a horizontal tough direction is always straight and horizontal. This is the shortest path a crack can take in this sample. For vertically tough samples, the fracture path becomes more and more complex as fracture anisotropy increases. Indeed, the crack path is straight for the lowest anisotropy (left column in Fig.~\hyperref[fig:2]{\ref{fig:2}c}), but becomes longer for higher anisotropy, see also Supplementary Movie S2. It actually spans the entire sample for the highest value of fracture anisotropy (right column of the figure). As expected, aligning the strong unit cell direction with that of the loading makes the metamaterial tougher (Fig.~\hyperref[fig:2]{\ref{fig:2}d}). 
This confirms that the stress and strain at break are always higher for vertical samples than for horizontal samples. 

Those higher stresses at break coincide with longer crack and are directly controlled by anisotropy. Both the length and widths of the crack increase with the unit cell anisotropy (Fig.~\hyperref[fig:3]{\ref{fig:3}a}). 
The crack length stays equal to the sample width for horizontally-tough samples, but increases rapidly with the fracture anisotropy for vertically tough samples. This is because the higher the anisotropy of the unit cell, the more stresses are concentrated in the vertical direction (Fig.~\hyperref[fig:2]{\ref{fig:2}b}), i.e., the crack is more likely not to follow the horizontal direction.

\begin{figure*}[t!]
 \begin{center}
  \includegraphics[width=2\columnwidth,clip,trim=0cm 0cm 0cm 0cm]{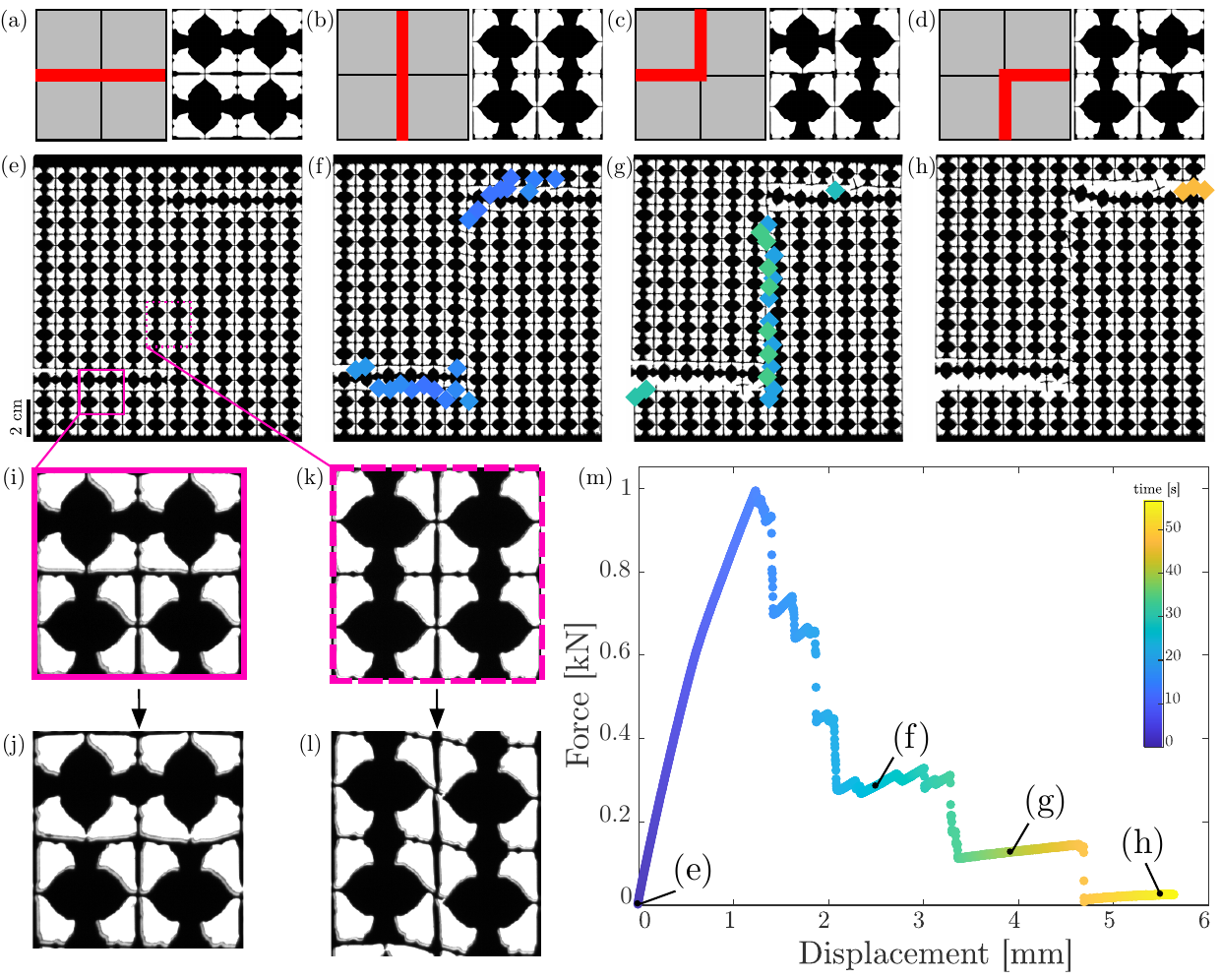}
 \end{center}
 \caption{(a-d) Design principle for crack steering in horizontal and vertical directions, as well as $\SI{90}{\degree}$ turns. The unit cells used here have a fracture anisotropy ratio $J_1/J_2 = 15$.
 (e-h) Image sequence showing the fracture of a $12 \times 12$ sample with a pre-programmed crack path in the shape of the letter ``S''. Images are taken at times \SI{0}{\second}, \SI{23}{\second}, \SI{36.9}{\second}, and \SI{54.2}{\second}, which correspond to a global strain of 0, 0.016, 0.025 and 0.038, respectively. The disks denote the fracture locations in the unit cells and the colors of the disks encode time (colorbar in (m)). See also Supplementary Movie S4. 
 (i-j) Zoom-in on figures (e) and (f), showing four unit cells located either side of the horizontal part of the crack path. 
(k-l) Zoom-in on figure (e) and (g), showing four vertical cells located either side of the central part of the crack path. 
 (m) Force-displacement curve corresponding to the experiment shown in (e-h).}
 \label{fig:5}
\end{figure*}

Intuitively, longer cracks release more fracture energy since they break more molecular bonds and indeed, we see that the energy dissipated by the crack increases with the fracture anisotropy ratio for vertically-tough samples, but decreases for horizontally-tough samples (Fig.~\hyperref[fig:3]{\ref{fig:3}b}). This is because we maximize the fracture anisotropy ratio not only by making one direction tougher, but also by making the other direction weaker. For the most anisotropic sample, this leads to a 100-fold difference in the dissipated energy between the vertically  and horizontally-tough samples. Interestingly, the stiffness of the samples barely varies for all samples, as shown in Fig.~\hyperref[fig:3]{\ref{fig:3}c}, and is around \SI{290(77)}{\mega\pascal}. We interpret this fact by two compensating effects during the elastic loading phase. In both horizontal and vertical orientations, the stress is concentrated in the ligaments parallel to the loading direction. While those ligaments are thinner for the horizontal samples, they are also shorter than the thicker ligaments of the vertical samples. These findings suggest two important conclusions: First, anisotropy in energy release rate can be used to steer cracks in heterogeneous structures, and cracks tend to propagate along the weak direction of the unit cells, \emph{e.g.}, vertically in samples shown in Fig.~\hyperref[fig:2]{\ref{fig:2}c}. Second, anisotropy allows us to control the energy dissipated by the crack without compromising the stiffness of the structure. Achieving tortuous fracture paths is a particularly interesting approach for applications where both a specific energy absorption and structural stiffness are required. 

\section*{Maximising crack length}

\begin{figure*}[t!]
 \begin{center}
  \includegraphics[width=2\columnwidth,clip,trim=0cm 0cm 0cm 0cm]{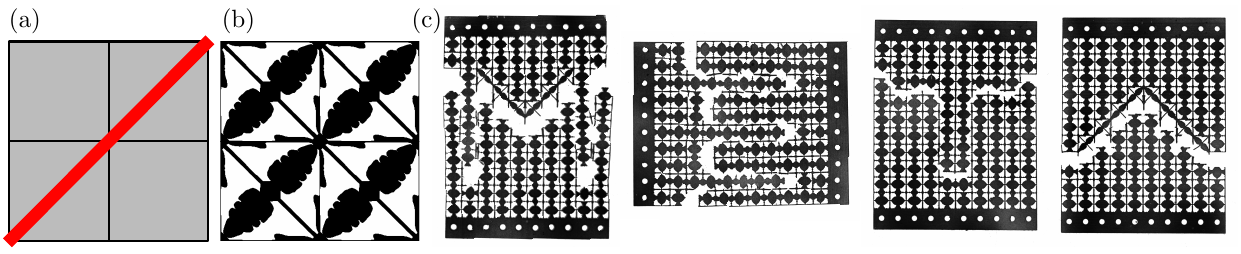}
 \end{center}
 \caption{\textbf{On-demand fracture.} (a-b) Design principle for crack steering in the diagonal direction. 
 (c) Specimens made with 10 $\times$ 10 unit cells after a unidirectional tension in the vertical direction, that break following a path that outlines the letters ``M'', ``E'', ``T'' and `` A''. Note that the letter ``E'' has been rotated by 90 degrees. Unit cells are 12 mm wide. 
 }
 \label{fig:6}
\end{figure*}

In the preceding section we investigated unit cells with different fracture anisotropy ratio. Here we investigate how different crack path designs can influence the total fracture length to obtain a maximum value---with its corresponding maximum dissipated energy. To that end, and inspired by the tortuous geometry of the crack in Fig.~\hyperref[fig:2]{\ref{fig:2}c} (rightmost column), we arrange the unit cells in a metamaterial in order to create comb-like paths. Starting from a configuration that contains only vertically tough cells ($n_x \times n_y$ cells), we replace some of them with horizontally-tough ones in a staggered order (see schematic in Fig.~\hyperref[fig:4]{\ref{fig:4}b}), thereby creating a weak path for the crack to follow; these \textit{weak cells} are then separated horizontally by a single unit cell and vertically by $\ell$ unit cells. We perform experiments with an increasing number of $\ell$ to obtain the longest comb-like crack path, as we show in Fig.~\hyperref[fig:4]{\ref{fig:4}a}and in Supplementary Movie S3.

From left to right, the figure shows increasing values of $\ell$, from $\ell = 3$ (figure labeled \textit{i}) to $\ell = 15$ (figure labeled \textit{vi}). As apparent from the figure, the crack follows the prescribed paths, and the vertical ``teeth'' of combs are one unit cell wide. For higher comb amplitudes ($\ell>\ell_{max} = 8$), the crack takes horizontal shortcuts (see, for instance, the orange arrow in the figure labelled \textit{iv}). This makes the fracture path shorter than what we expect. We repeat the experiment for other unit cells, and measure crack length (in number of broken unit cells) as a function of $\ell$ (Fig.~\hyperref[fig:4]{\ref{fig:4}c}), for all four unit cell designs (keeping the same colour code as in previous figures). For low values of $\ell$, we see a linear increase of the measured crack path with $\ell$ for all unit cells, apart for the least anisotropic one (in green). This linear increase follows the geometric law $\ell \times (n_y-1) + n_x$ (shown as a dashed line), which simply confirms that the crack follows the target path. The experimental data points depart from this prediction for values of $\ell$ larger than a certain threshold $\ell_{\max}$, whose value depends on the fracture anisotropy ratio of the unit cell. As expected, the larger this ratio, the larger is the value of $\ell_{\max}$ at which the designed path is no longer followed; this value is, respectively, equal to 1, 6, and 11, for the unit cells with an anisotropy ratio 1.7, 4.5, and 15. For the most anisotropic unit cell (in red) the samples always broke along the path, even for the largest sample we could manufacture (with $\ell=33$).

Fig.~\hyperref[fig:4]{\ref{fig:4}d} illustrates the maximum comb amplitude $\ell_{max}$ measured experimentally as a function of the fracture anisotropy ratio of the unit cell. For the most anisotropic unit cell, we could not measure $\ell_{max}$ (indicated in the figure with a vertical arrow). The linear trend observed in the figure suggests a direct correlation between the anisotropy ratio and the maximum comb amplitude obtained. Breaking a vertical structure of the target path (delimited by three red cells in Fig.~\hyperref[fig:4]{\ref{fig:4}b}) requires to break $\sim \ell$ weak unit cells, while taking a shortcut requires to break one unit cell in the strong direction. For $\ell \gtrsim J_1/J_2$ it becomes energetically favourable to take a shortcut, and break a horizontally strong unit cell instead of several vertically weak cells. The dashed line in Fig.~\hyperref[fig:4]{\ref{fig:4}d} corresponds to $\ell_{max} = J_1/J_2$, which agrees reasonably well with our experimental data.

\section*{Fracture mechanism}

But what mechanism could lead to fracture that is parallel to the loading path? We now design samples that break following more complex paths. We consider paths that span the entire width of the sample, that will break the sample into two (top and bottom) parts. We outline the desired path on a square grid (e.g., the red lines in Fig.~\hyperref[fig:1]{\ref{fig:1}a}), and we then use the design principle in Figs.~\ref{fig:5}a-d to steer the cracks: we align locally the weak directions of the unit cells with the path. This allows us to steer the crack either horizontally (Fig.~\hyperref[fig:5]{\ref{fig:5}a}), vertically (Fig.~\hyperref[fig:5]{\ref{fig:5}b}) or at an angle (Fig.~\hyperref[fig:5]{\ref{fig:5}cd}).
To explain the mechanism behind crack deviation, we design a material that breaks with the shape of a step function, and that we show in Fig.~\hyperref[fig:5]{\ref{fig:5}e}. This exemplifies a situation where one might need to deviate a crack away from a critical area.

In Fig.~\hyperref[fig:5]{\ref{fig:5}f}, we see that the crack first propagates horizontally along the bottom left and then along the top right of the sample (highlighted in blue), see also Supplementary Movie S4. There, mode I fracture predominates, as we show in Fig.~\hyperref[fig:5]{\ref{fig:5}ij}: The horizontal cells break at very low strains (locally we estimate that $\varepsilon \simeq 0.03$ from Fig.~\hyperref[fig:5]{\ref{fig:5}i} and \hyperref[fig:5]{\ref{fig:5}j}) and with very little deformation of the whole specimen. This instance corresponds to the first large drops in the force (for a global strain of about $0.01$ in Fig.~\hyperref[fig:5]{\ref{fig:5}m}). After these two horizontal sections break (in blue in Fig.~\hyperref[fig:5]{\ref{fig:5}f}), the vertical central part is strongly loaded in mode II. This leads to the shearing of the vertical cells, which can locally accommodate large strains without failure ($\gamma \simeq 0.13$ from Fig.~\hyperref[fig:5]{\ref{fig:5}k} and \hyperref[fig:5]{\ref{fig:5}l}). This corresponds to the second large drop in load curve, for a global strain of about $0.023$ in Fig.~\hyperref[fig:5]{\ref{fig:5}m}). The last unit cells to break (highlighted in yellow in Fig.~\hyperref[fig:5]{\ref{fig:5}h}) are located at the very top right, and we impute this to the fact that the loading is not perfectly uniform over the whole specimen. We attribute the strong anisotropy in the fracture response to the slender nature of the beams that compose the unit cells. The beams within the unit cell bend when the latter is sheared, allowing for larger deformations that on tension.

\begin{figure}[h!]
 \begin{center}
  \includegraphics[width=1\columnwidth,clip,trim=0cm 0cm 0cm 0cm]{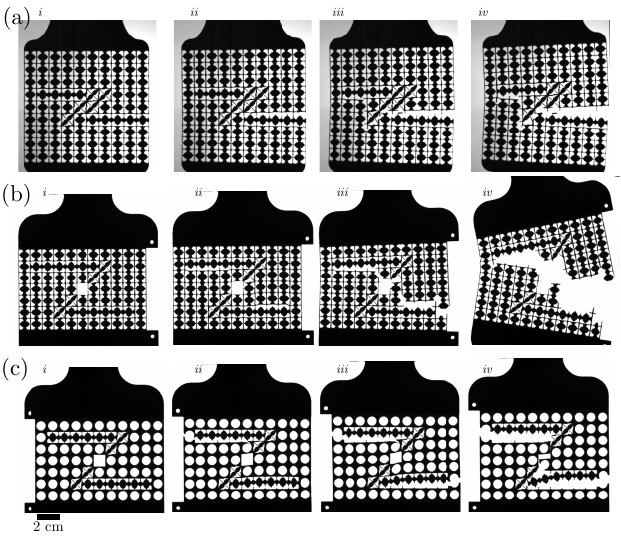}
 \end{center}
 \caption{\textbf{Reverse fracture.} (a) and (b) Failed attempts at a Z crack, using unit cells with anisotropy in energy release rates of 15 as well as diagonal cells (see Materials and Methods for details). In (a) the crack is initiated at the right side of the specimen, in (b) it is initiated in the middle by removing a unit cell there. 
 (c) Z shaped crack obtained in a specimen made of circular unit cells, horizontal unit cells with anisotropy 15 and diagonal unit cells, see also Supplementary Movie S5.
 }
 \label{fig:7}
\end{figure}

In simple terms, the fracture mechanism at play here is the following: in the metamaterial under load, the first slender elements to break are the ones loaded under tension---\emph{viz.} they are parallel to the loading direction. As a result of this failure, the stress within the metamaterial redistributes. Because of this stress redistribution, the slender elements that are perpendicular to the loading direction will be loaded under shear. Since they are slender, they can bend and thereby accommodate a large strain before breaking and hence appear as ductile.

Eventually though, those slender elements loaded under shear will break. This sequential failure enabled by the bending mechanics of slender element and redistribution of stress is key to achieving cracks of arbitrary complexity.

\section*{Designer cracks}

In this example of a step-shaped fracture, we see that crack steering is achieved through a transition in fracture modes, \emph{i.e.}, from mode I to mode II enabled by bending. 

As a consequence, the parts of the crack path in tension (mode I) break before the parts that are in shear (mode II), which ultimately yields the desired crack path.
This strategy works well in most cases, and the designs studied so far can be used to set only horizontal and vertical cracks, and a combination thereof. To improve the sharpness of our designs and to expand the range of patterns we can break, we add one more cell that was designed so that, in the same square computational domain, the fracture anisotropy is maximized at a \SI{45}{\degree} angle (see \hyperref[sec:methods]{Materials and Methods}).

With this new unit cell, shown in Fig.~\hyperref[fig:6]{\ref{fig:6}a-b}, we can program different crack paths following the same principle outlined above with the horizontally and vertically tough unit cells. We used all these unit cells to design and fabricate samples that outline the letters ``U'', ``V'' and ``A'' (Fig.~\hyperref[fig:1]{\ref{fig:1}c}) and also those that make up the word ``META'' (Fig.~\hyperref[fig:6]{\ref{fig:6}c}).

\subsection{Reverse crack}

We now turn our attention to even more complex crack paths, with a fracture that at some point reverses direction. We opted for a zigzag design, coincidentally described by the letter ``Z'' (see Fig.~\hyperref[fig:7]{\ref{fig:7}a-\textit{i}}). To program this crack, we first use the design principle in Figs.~\ref{fig:5} and \ref{fig:6}, where we start with a specimen made almost exclusively of vertically strong unit cells, apart from two horizontal lines and a diagonal line. We introduce a notch on the right of the sample to initiate the crack there. Once under tension, we see that the crack propagates to the left, as expected (Fig.~\hyperref[fig:7]{\ref{fig:7}a-\textit{ii}}); however, as soon as it hits the diagonal cells, the crack does not follow our programmed path (Fig.~\hyperref[fig:7]{\ref{fig:7}a-\textit{iii}}). Ultimately, the crack front reaches the second line of horizontally-tough cells and the sample is broken into two parts, without showing the expected programmed shape (Fig.~\hyperref[fig:7]{\ref{fig:7}a-\textit{iv}}). 

We attribute this effect to a very large stress intensity factor in mode I once the fracture front reaches the diagonally-tough unit cells (Figs.~\hyperref[fig:7]{\ref{fig:7}a-\textit{ii}}), because at this stage the crack is rather long. To preclude this effect, we decide to nucleate the crack in the middle of the sample, creating a notch by removing the central unit cell (see Fig.~\hyperref[fig:7]{\ref{fig:7}b}); we also replace the horizontal unit cells on the first and last columns with vertically tough cells, so that the crack does not nucleate on the edges. Once under tension, we observe that the first unit cells that break are the horizontally-tough ones located at either side of the diagonal line, since these are easier to break. Next the vertical cells in the central part break but not the diagonal cells (Fig.~\hyperref[fig:7]{\ref{fig:7}b-\textit{iii}}). Instead, the crack has taken a shortcut, and breaks the sample into two, creating a shape that still does not resemble the letter ``Z'' (see Fig.~\hyperref[fig:7]{\ref{fig:7}b-\textit{iii}}).

The vertical cells located around the central part are too easy to break in mode I, which explains why the crack tip takes a horizontal shortcut. We thus decide to take our design one step further, and to replace the vertically tough unit cells with those that have the lowest energy release rate when subjected to biaxial loading: a square cell with a circular hole~\cite{zhang_tailoring_2022}. The specimen obtained is shown in Fig.~\hyperref[fig:7]{\ref{fig:7}c-\textit{i}}) and it consists mostly of circular unit cells with the same volume fraction of material as the topologically optimised ones (i.e., 50\%)---in addition to horizontally- and diagonally-tough cells that delineate the letter ``Z''. Once under tension, the horizontal cells break first as before (see Fig.~\hyperref[fig:7]{\ref{fig:7}c-\textit{ii}}), but then the crack propagates to the center (Fig.~\hyperref[fig:7]{\ref{fig:7}c-\textit{iii}}) and then diagonally as originally programmed. The sample is finally broken into two halves with the intended shape. This path with extreme turning angles is a powerful illustration of the potentiality of our design principle for crack steering.

\section*{Conclusion}

We have shown that metamaterials based on topology optimised unit cells open promising routes for fracture control. Their versatility stems from the slender struts that compose the unit cells, making it break readily in tension but bend significantly in shear. The shape of the crack can therefore be programmed into the material, even for intricate paths.

Unlike other strategies that use complex microfabrication techniques or complex loading sequences, we design a metamaterial whose fracture behaviour is solely dictated by its geometry. This type of metamaterial is thus easy to create once the unit cell designs are known, and we posit that this principle generalizes to other base materials and other scales, as well as other loading scenarios (e.g., compression, bending, twisting). The combination of bending and stretching had been already used in the context of double-network elastomers with enhanced toughness~\cite{Sun_Nature2012, deng_nonlocal_2023} so a natural question would be to ask whether cracks of arbitrary complexity could also be achieved in such materials.

Further research is however required to extend our strategy to ductile materials, as we used linear elastic fracture mechanics and brittle base materials. Similarly, multi-phase metamaterials would be promising to study, which would imply solving for solid-solid interfaces within the unit cell. More importantly, while in 2D the crack design seems somewhat straightforward (we can rely on intuition), the true potential of these design principles will be unleashed when we explore to 3D fracture. 

 For these future lines of research we deem necessary to understand the underlying mechanisms that governs the extreme fracture behaviours we observe: Is the bending  of the thin trusses essential for crack steering? Or could the same level of anisotropy and crack steering capacity be obtained in simpler unit cells design (\emph{i.e.} without thick and thin trusses)?  It will also be interesting to assess how well these metamaterials perform in practical cases, such as strong impacts rather than quasi-static loading. This will be the case in most applications where fracture is unwanted, such as mechanical fuses and impact mitigation, but also food or even everyday examples such as the shattering of one’s phone screen.

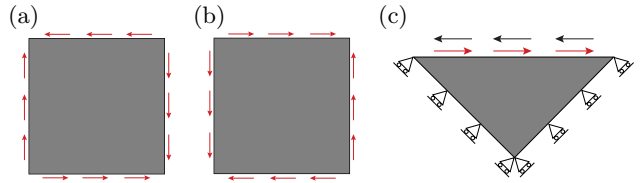
\begin{figure}[h!]
  \begin{center}
    \begin{picture}(240,60)
      \put(5,0){\includegraphics[height=2cm]{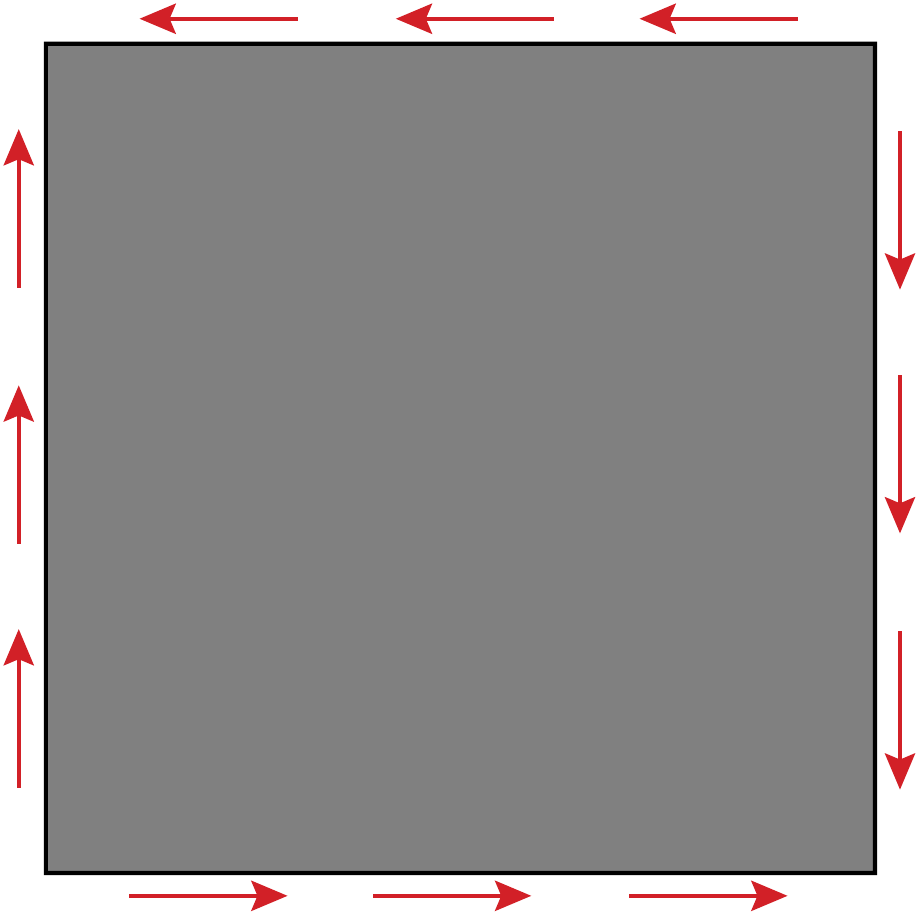}}
      \put(75,0){\includegraphics[height=2cm]{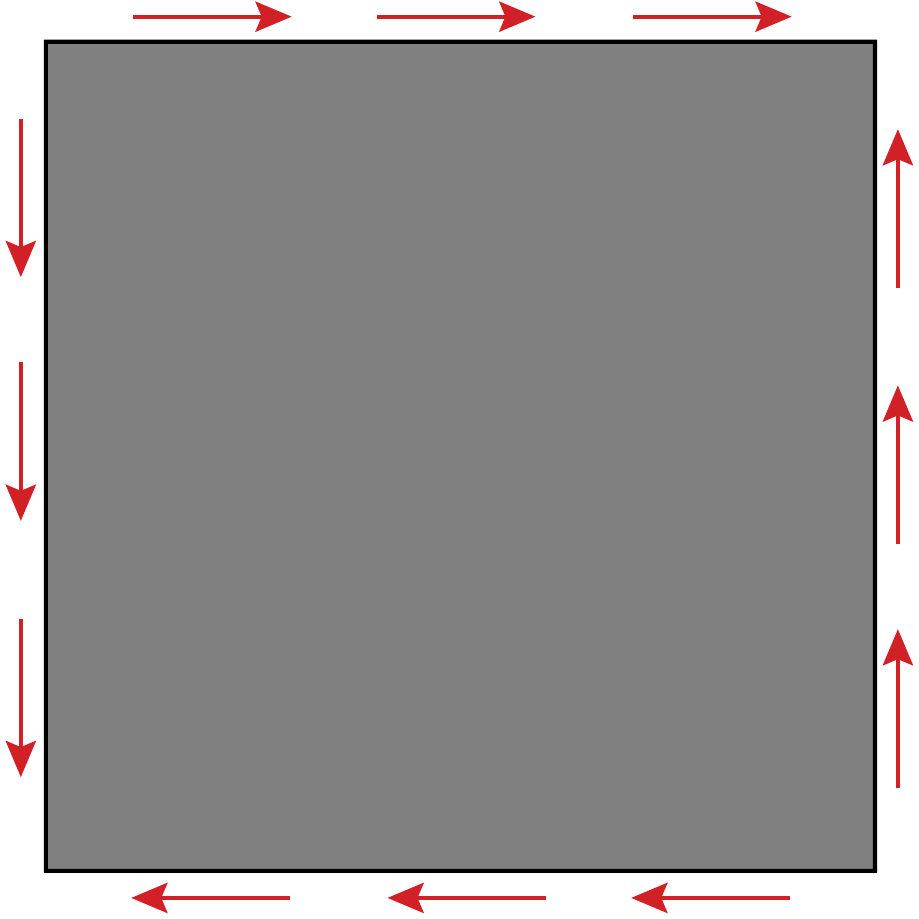}}
      \put(145,0){\includegraphics[height=2cm]{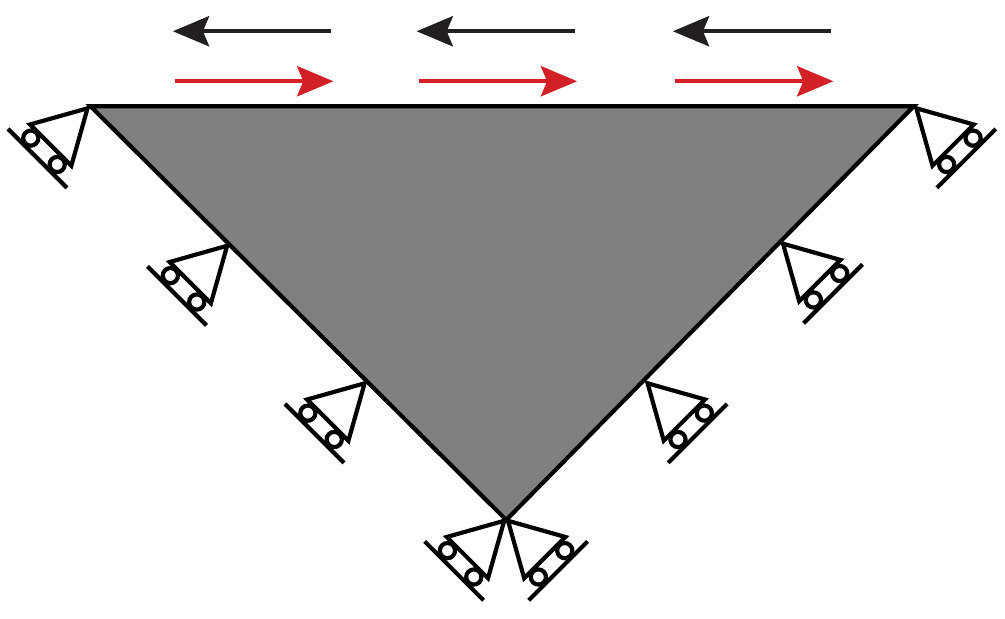}}
      \put(0,60){(a)}
      \put(70,60){(b)}
      \put(140,60){(c)}
    \end{picture}
  \end{center}
  \caption{Computational domain used to design the unit cell for shear loading: (a,b) Two shear loading cases in~\eqref{eq:optimization}; and (c) Reduced computational domain due to symmetry. }
  \label{fig:shear_model}
\end{figure}

\section*{Materials and methods}
\label{sec:methods}
\subsection*{Design of the unit cells}

The systematic design of a unit cell with highly anisotropic fracture behavior is conducted in this work by means of topology optimization (TO). In a nutshell, TO is an iterative procedure that combines finite element anaylisis (FEA) with a gradient-based optimizer. Specifically, at each iteration of the optimization, FEA is conducted to evaluate the objective function---in our case a quantification of fracture anisotropy---and sensitivity information is extracted and used by the optimizer to improve the design for the next iteration.
The zeroth value of a level set function is used to describe the topology, and the method of moving asymptotes (MMA) is used as the optimizer. 

In our work we build some concepts of linear elastic fracture mechanics (LEFM) into a topology optimization framework that uses an enriched finite element formulation to evaluate the objective function and \textit{topological derivatives} to evaluate energy release rates throughout the boundary of the unit cell from a single FEA.
For a given load case $i$, for instance compressing the unit horizontally, energy release values obtained are then aggregated into a single value
\begin{equation}
    J_i = \frac{1}{N} \sum_{j=1}^N G_j,
\end{equation}
where $G_j$ denotes the energy release rates that is evaluated at $N$ locations (enriched nodes) along the unit cell boundary.

Then the optimization problem is formally stated as
\begin{equation} \label{eq:optimization}
\begin{split}
\text{minimize }  & \Phi  = \omega J_1 - (1-\omega) J_2, \\
\text{such that } & \quad \boldsymbol K_1 \boldsymbol U_1 = \boldsymbol F_1, \\
& \quad \boldsymbol K_2 \boldsymbol U_2 = \boldsymbol F_2, \\
 & \quad V = \bar{V},
\end{split}
\end{equation}
where $V$ is the volume of solid material that is constrained to a target value $\bar{V}$ (for instance 50\% of the computatinal design domain), $\boldsymbol K_i \boldsymbol U_i = \boldsymbol F_i$ is the discrete system of equations corresponding to the $i$th loading case (either vertical or horizontal), $\boldsymbol{K}_i$ is the corresponding stiffness matrix,  $\boldsymbol{F}_i$ is the force vector, and $\boldsymbol{U}_i$ is the degree of freedom vector. In \eqref{eq:optimization} $\omega$ is a factor that is used to give priority to one loading case over the other, so for $\omega = 0.5$ we give the same priority to both loading scenarios. 

The design of unit cells for vertical and horizontal loading has been thoroughly discussed elsewhere~\cite{zhang_tailoring_2022}. Here we showcase the methodology for optimizing the unit cells that maximize fracture anisotropy when subjected to shear loading. Fig.~\ref{fig:shear_model} shows the loading cases in~\eqref{eq:optimization} (Figs.~\hyperref[fig:shear_model]{\ref{fig:shear_model}a-b}) and also the actual computational domain used during topology optimization (Figs~\hyperref[fig:shear_model]{\ref{fig:shear_model}c}), which is reduced by taking advantage of symmetry. The final design obtained after topology optimization can be visualized in Fig.~\hyperref[fig:6]{\ref{fig:6}b}, where it can be seen that unit cells are tougher along \SI{45}{\degree} and weaker along \SI{135}{\degree}.

For more details on the procedure herein, the reader is referred to a thorough description by Souto et al.~\cite{zhang_tailoring_2022} and Zhang et al.~\cite{souto_edible_2022}.

\subsection*{Fabrication of the samples}

The samples were 3D printed with a polyjet 3D printer (Connex 500, Stratasys), which has a precision on the order of 200 $\mu$m. Taking this limit into account, the unit cells are scaled up to a size 12 mm to ensure a satisfying rendering of thinnest parts the unit cells. We used a brittle polymer (Vero, Stratasys) and printed the samples with a thickness 3mm. The largest sample we could print was 33 unit cells long, which was the maximum the print bed could hold. The printer also induces anisotropy due to the fabrication process, so we made sure to always print samples in the same direction (see also Supplementary Information).

\subsection*{Testing}

The 3D printed samples were then fastened in custom-made clamps and then placed in a testing machine (Instron 5985) equipped with a 300 kN load cell, which enabled us to impose a tensile displacement with an accuracy of 0.01 mm and to record the force with $0.5\%$ of the reading accuracy. The sample was then put under tension at an imposed displacement rate of 0.1mm/s, and the experiment was recorded using a USB camera (ac2040, Basler with a 50 mm lens) at 25 FPS.

\subsection*{Simulations in COMSOL}

We use the 2D solid mechanics module from COMSOL. We model a linear elastic material and we impose an infinitesimal displacement of $\varepsilon =  0.008$ at the edge of a 10$\times$10 specimen, with a very long notch that spans 5 unit cells and that is a quarter of unit cell in thickness. This method does not allow us to predict fracture propagation per se, which would actually be rather challenging given that the crack needs to nucleate numerous times to propagate. As a proxy, we identify the loci of maximum von Mises stress, which corresponds to the direction in which the crack is most likely to propagate. Note that in Fig.~\hyperref[fig:2]{\ref{fig:2}b} we present the 4 central cells, located around the tip of the notch. 

\begin{acknowledgments}

The authors would like to thank C. Ederveen Janssen, D. Giesen, S. Koot, M. van Hecke and D. Ursem for their technical help. We acknowledge funding from the European Research Council under grant agreement 852587 and the Netherlands Organisation for Scientific Research under grant agreement NWO TTW 17883.
All the codes and data supporting this study are available on the public repository \url{https://doi.org/10.5281/zenodo.11103085}.
\end{acknowledgments}

\bibliography{Fracture}

\end{document}